\documentclass{Rinton-P9x6}

\begin{document}

\title{SEARCH FOR \lq\lq LARGE" EXTRA DIMENSIONS AT
THE TEVATRON}

\author{T. Ferbel}

\address{ Department of Physics and Astronomy\\
University of Rochester\\ Rochester, NY 14627}




\maketitle

\abstracts{ We report on a search for extra spatial dimensions
compactified at radii that are vast compared to those of the
Planck length, and even larger than the distance corresponding
to the scale of electroweak symmetry breaking. The study is based
on the suggestion that there may be only a single mass scale
appropriate to particle phenomena, this being the effective Planck
scale $M_S$, which also serves as the scale for electroweak symmetry
breaking.  The only results that have been presented thus far from
the Tevatron are based on data for $e^+e^-$ and $\gamma\gamma$
production at D\O, and provide the most restrictive lower limits
to date on $M_S$ of $\approx$ 1 TeV. }

\section{Introduction}

Many particle physicists have become convinced that the
unification of particle interactions will require the presence of
a total of ten spatial dimensions, three of which are within the
reach of our senses and the others compactified at inconceivably
small radii corresponding to the Planck length ($R_P$) of $\hbar/M_Pc
\approx 10^{-33}$ cm, where $(M_P)$ is the Planck mass. Because it
is unlikely that our governments will ever provide accelerators
that can challenge the Planckian regime, it was therefore exciting
to learn of the possibility of sensing the impact of strings and
quantum gravity at far lower energies.

It is commonly accepted that the Standard Model (SM) is a
low-energy approximation to a more complete theory, the impact of which
should become apparent at a scale of the order of the Higgs mass,
i.e., $\approx$ 1 TeV.  The two most popular candidates for such a
theory have been based on supersymmetry and on strong dynamics.
More recently, Arkani-Hamed, Dimopoulos and Dvali (ADD),\cite{Arkani}
suggested that there may be only a single scale in particle phenomena,
that being the electroweak scale, which also corresponds to the
effective Planck scale $M_S\approx$ 1 TeV, provided that there exist
extra dimensions compactified at far larger radii than the
Planck length. This exciting idea has inspired experimenters and
phenomenologists to search for such large extra spatial dimensions,
and for the effect of quantum gravity in particle interactions.

The basic notion requires the particles of the SM to be localized to
our four-dimensional world, but allows gravitons to propagate in all
the large extra dimensions. In the presence of $n$ such dimensions, the
potential in Newton$'$s Law changes from its characteristic $r^{-1}$
dependence to a steeper form:

$$V(r) = \frac{\hbar c}{M_P^2} \ \frac{M_1M_2}{r} \longrightarrow
\frac{\hbar c}{(M_P^*)^{n+2}} \ \frac{M_1M_2}{r^{n+1}}$$

\noindent where $M_P^*$ is the Planck mass corresponding to
$(n+3)$ spatial dimensions, and Newton$'$s gravitational constant is
$G_N = \hbar c/M_P^2$.  Clearly, the lack of departure from
$r^{-1}$ behavior for distances greater than $\approx$ 1 cm rules out
the possibility that any of these extra dimensions are of macroscopic
size.  But for $n$ extra dimensions, compactified on sufficiently
small (and same size) radii $R$, the gravitational energy for
$r>>R$ retrieves its acceptable $r^{-1}$ dependence, with $M_P^2
\approx (M_P^*)^{n+2} (Rc^2/\hbar c)^n$.  Hence, choosing $R>>R_P$,
can effectively reduce $M_P^*$, and thereby increase the strength of
the gravitational coupling to $\hbar c/M_P^{*2}$ for the larger
compactified dimensions.  If $M_P^*$ is taken as the electroweak
scale $M_S \approx$ 1 TeV, then its value becomes coupled to the
compactification radius and to the number of such compactified
dimensions of size $R = (\hbar/2\sqrt{\pi} M_Sc)\
(M_P/M_S)^{2/n}$, where the extra factor $2\sqrt{\pi}$ is a
consequence of applying Gauss$'$s Law on an $n$-torus.\cite{Arkani}

Thus, for $n = 1, R \approx 8 \times 10^{14}$ cm; for $n =
2, R \approx 0.07$ cm; and, for $n = 3, R\approx 3 \times 10^{-7}$
cm.  On the basis of the agreement of Newton$'$s law with the
$r^{-1}$ dependence for gravitational interactions at macroscopic
distances, we can exclude $n = 1$.  Using cosmological arguments
(and recent measurements of the gravitational force at distances of
the order of 0.02 cm~\cite{Adelberger}), ADD exclude a scale as small as
$\approx$ 1--10 TeV for only two compactified dimensions.  However,
scales in this range cannot be excluded for 3 or more compactified
dimensions, and they can be probed only through elementary particle
interactions at high energies, where gravity is enhanced.

Compactified dimensions greatly increase the effective strength of
gravitational interactions through Kaluza-Klein $(G_{KK})$
modes of gravitons.\cite{Kaluza}  From the perspective of our (3+1) space-time
dimensions, the $G_{KK}$ modes correspond to massive gravitons,
with excitations spaced at $\hbar c/R$.  Each mode couples with
strength $G_N$, but, because of the large number of modes at
high energies, gravitational coupling is greatly increased (to the
order of $\hbar c/M_S^2)$.

Kaluza-Klein gravitons couple to the energy-momentum tensor, and
signatures for large extra dimensions depend on whether the
$G_{KK}$ in particle interactions are real (emitted in collisions)
or whether they are virtual (and do not leave our SM world).  Thus,
the impact of virtual $G_{KK}$ can be observed in reactions such
as $q\overline{q} \to G_{KK} \to \gamma\gamma$ , or $gg \to G_{KK}
\to e^+e^-$, etc.  $G_{KK}$ emission can lead to apparent
violation of energy and momentum (as well as of angular momentum)
conservation when the graviton has components in the bulk that
escape the brane of the SM, e.g., $q \overline{q} \to G_{KK} + g$,
or $e^+e^- \to G_{KK} + \gamma$, etc.  The characteristic
signatures for contributions from virtual $G_{KK}$ correspond to
yields of massive systems beyond rates expected on the basis of
the SM,  while direct emission of $G_{KK}$ results in an increase
of the yield of events with large apparent imbalance in transverse
momentum (or \lq\lq missing $E_T"$, $\not\!\!{E_T}$), in particular,
mono-jet events.

Limits on $M_S$ of $\approx$ 1 TeV have already been reported from
LEP,\cite{LEP} and somewhat weaker limits (from searches for virtual
graviton contributions) have been published by
experiments at HERA.\cite{HERA}

\section{Virtual Graviton Effects}

In this paper, we focus on virtual $G_{KK}$ effects,
primarily because the analyses of direct emission have not as yet
been completed at the Tevatron.  Also, since the results on
virtual $G_{KK}$ emission from CDF will not become available
before Spring 2001,\cite{Gerdes} we report only on the work from
D\O.\cite{PRL}

For the case of virtual graviton contributions, the amplitude for
$G_{KK}$ exchange has to be added coherently to that from the SM,
because processes such as $q\overline{q} \to \gamma^* \to e^+e^-$
and $q\overline{q} \to G_{KK} \to e^+e^-$ can provide important
interference terms.

Three phenomenological formulations of the problem have appeared in
the literature.\cite{Hewett,Giudice,Han} These are equivalent, and differ
only in their definitions of $M_S$.  D\O\ follows the more sophisticated
phenomenology,\cite{Han} which contains a dependence of the cross section
on $n$. The correspondence between the three definitions of scale is that
$M_S(n=5) \approx M_S(\lambda=+1)$ of Hewett,\cite{Hewett} and
$M_S(n=4) \approx \Lambda_T$ of Giudice et al.\cite{Giudice}

\section{Analysis at D\O}

D\O\ bases its analysis on both di-electron and di-photon signals.
Since D\O\ did not have a central magnetic
field in Run-1, the analysis ignores electric charge, and the cross
section is therefore examined as a function of $M_{ee}$ or $M_{\gamma\gamma}$,
and $|\cos\theta^*|$, where $\theta^*$ is the angle of the $e$ or $\gamma$
relative to the line of flight of the $ee$ or $\gamma\gamma$ system
(helicity frame). The analysis follows the Cheung-Landsberg
extension\cite{CheungL} of
previous studies that were based on the use of only the mass
variable in the problem.

Because the instrumental background (e.g., a jet mimicking a
photon or an electron) in this search is small at large
$E_T$,\cite{GregM} the signals for $ee$ and $\gamma\gamma$ are
added together in the comparison of data to theory. This maximizes
the reach of the experiment to highest mass scales because it
allows a loosening of the usual strict electromagnetic (EM) shower
requirements on both photons and electrons. The analysis ignores
charged-particle tracking information in the detector, and relies
purely on the observation of di-EM systems of high invariant mass
($M_{_{EM,EM}}$). The theory used to describe contributions from
the SM and virtual gravitons is a leading-order (LO) calculation,
and therefore is not affected by the summing of the $ee$ and
$\gamma\gamma$ components.\cite{CheungL} Nevertheless, the
expected yields are corrected for higher-order effects through an
application of a common ``K-factor" of 1.3 (this probably
underestimates the expected yield from gravitons at largest
($M_{_{EM,EM}}$). The criteria for the final 1250 candidate events
require only two acceptable EM showers, each with $E_T > 45$ GeV,
and no missing transverse momentum ($\not\!\!{E_T} < 25$ GeV).
There are no requirements placed on jets. In addition, there is a
less restrictive sample of di-EM events, that requires $E_T > 25$
GeV, which is used to check EM efficiencies through a comparison
to $Z \to ee$ events.

The data are compared to the LO parton generator of Cheung and
Landsberg,\cite{CheungL} augmented with a parametrized D\O\ detector
simulation package that models the acceptance, resolution, vertex smearing,
and the impact of having additional vertexes from overlapping multiple
interactions. As indicated above, the calculation applies a uniform K-factor
correction to all SM and virtual-graviton cross sections, and even
introduces a transverse impact to the di-EM system (assuming that it
is the same as in the inclusive $Z$ data of D\O~\cite{Dylan}). The
corrections used to account for higher-order (incident gluon Bremsstrahlung)
effects are relatively unimportant. The $CTEQ4LO$ parton distribution
functions ($PDF$) are used to integrate the matrix elements over the
incident parton distributions (with checks performed using other $PDF$s).

Most of the di-EM events arise from prompt di-photon, and, to a
lesser extent, from $e^+e^-$ production (usual SM processes). The
background from more exotic channels, such as $W+$jets, $W+\gamma, WW,
t \bar t$, etc, is miniscule. The largest instrumental background
($\approx 7\%$ of the di-EM signal) is from multijet production,
where two jets are misidentified as EM showers.

\section{Comparison of Data with Monte Carlo}

\begin{figure}
\centerline{ \epsfxsize=4.truein \epsfbox{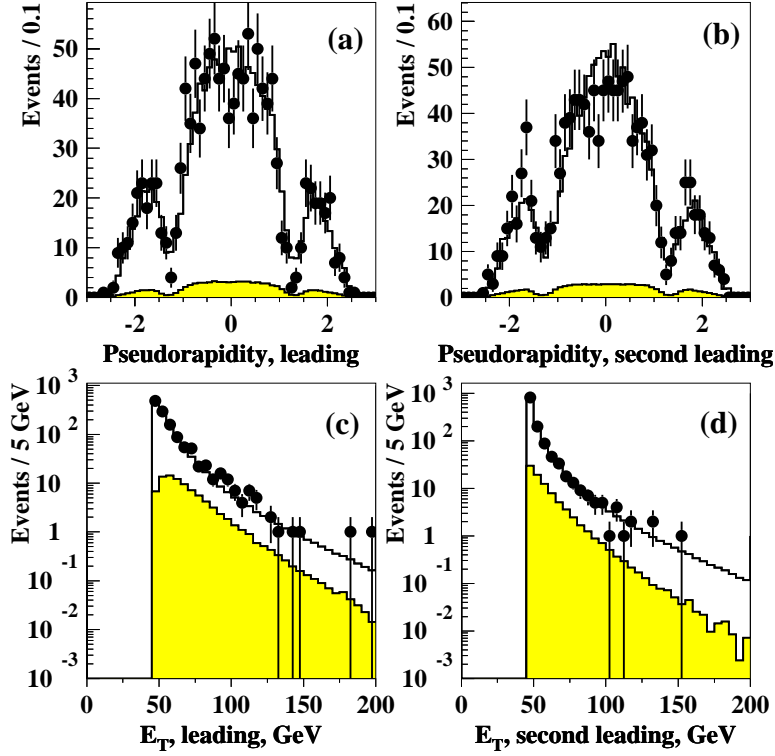}}
\caption{At the top is a comparison of the pseudo-rapidity
distribution for data with the Monte Carlo, and at the bottom of
$E_T$, for the two EM objects.} \label{fig:eta_MC}
\end{figure}

\begin{figure}
\centerline{ \epsfxsize=4.5truein \epsfbox{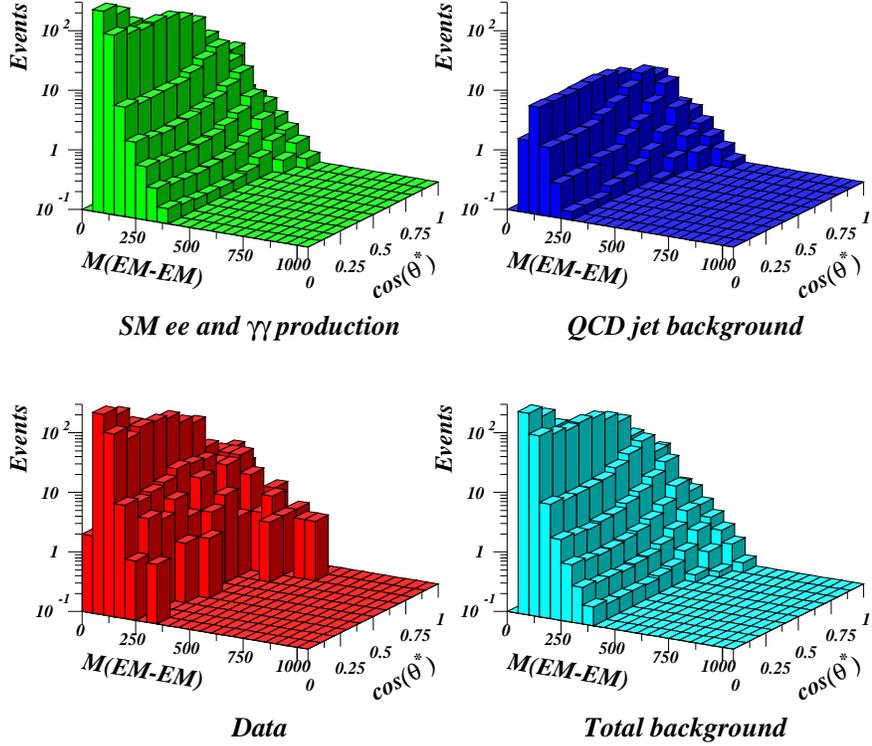}}
\caption{Comparison of data with Monte Carlo in the
($M_{_{EM,EM}},|\cos\theta^*|$) plane, for contributions from all
processes expected in the SM} \label{fig:two_dim}
\end{figure}

\begin{figure}
\centerline{ \epsfxsize=2.5truein \epsfbox{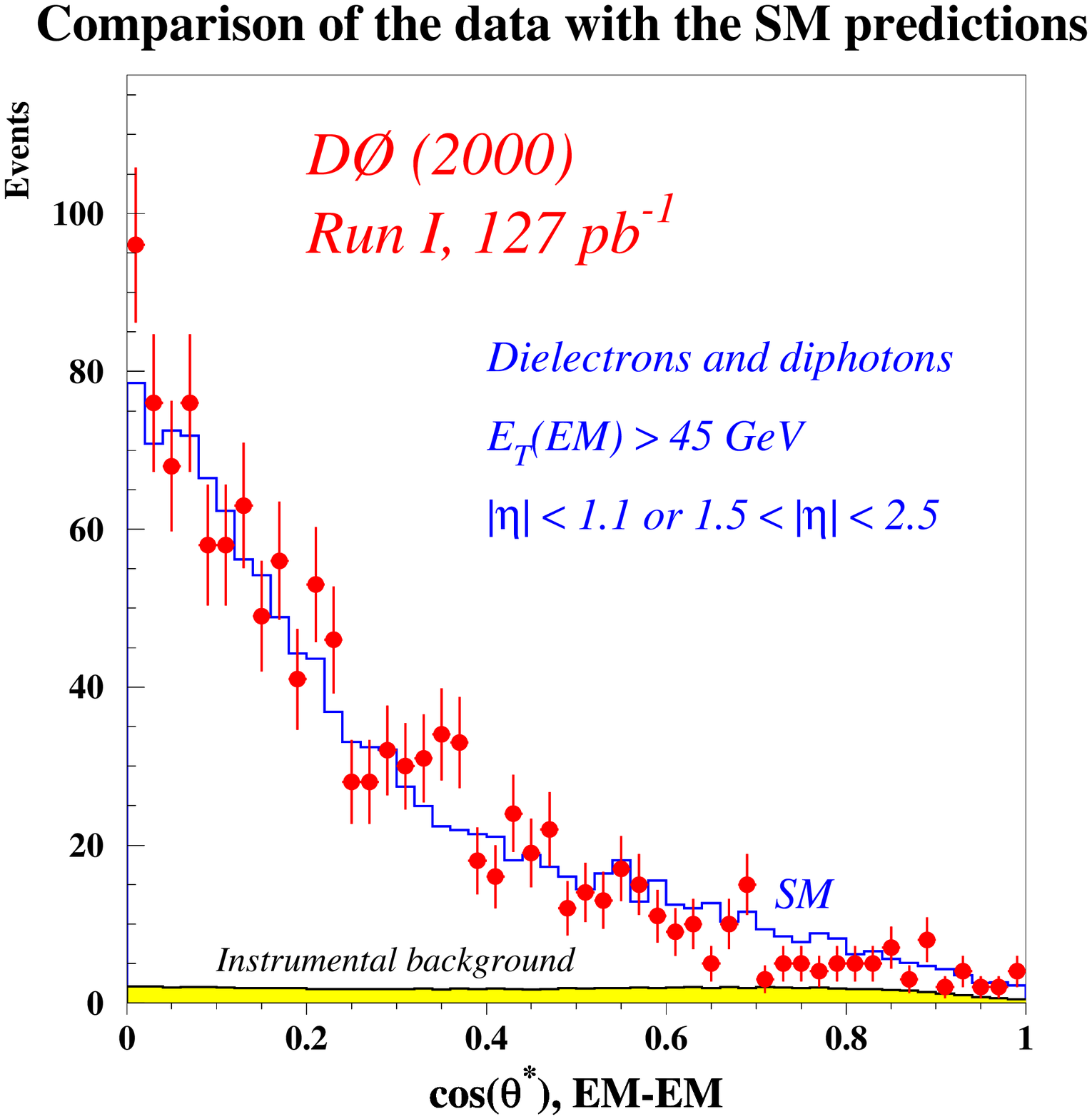}
\epsfxsize=2.5truein \epsfbox{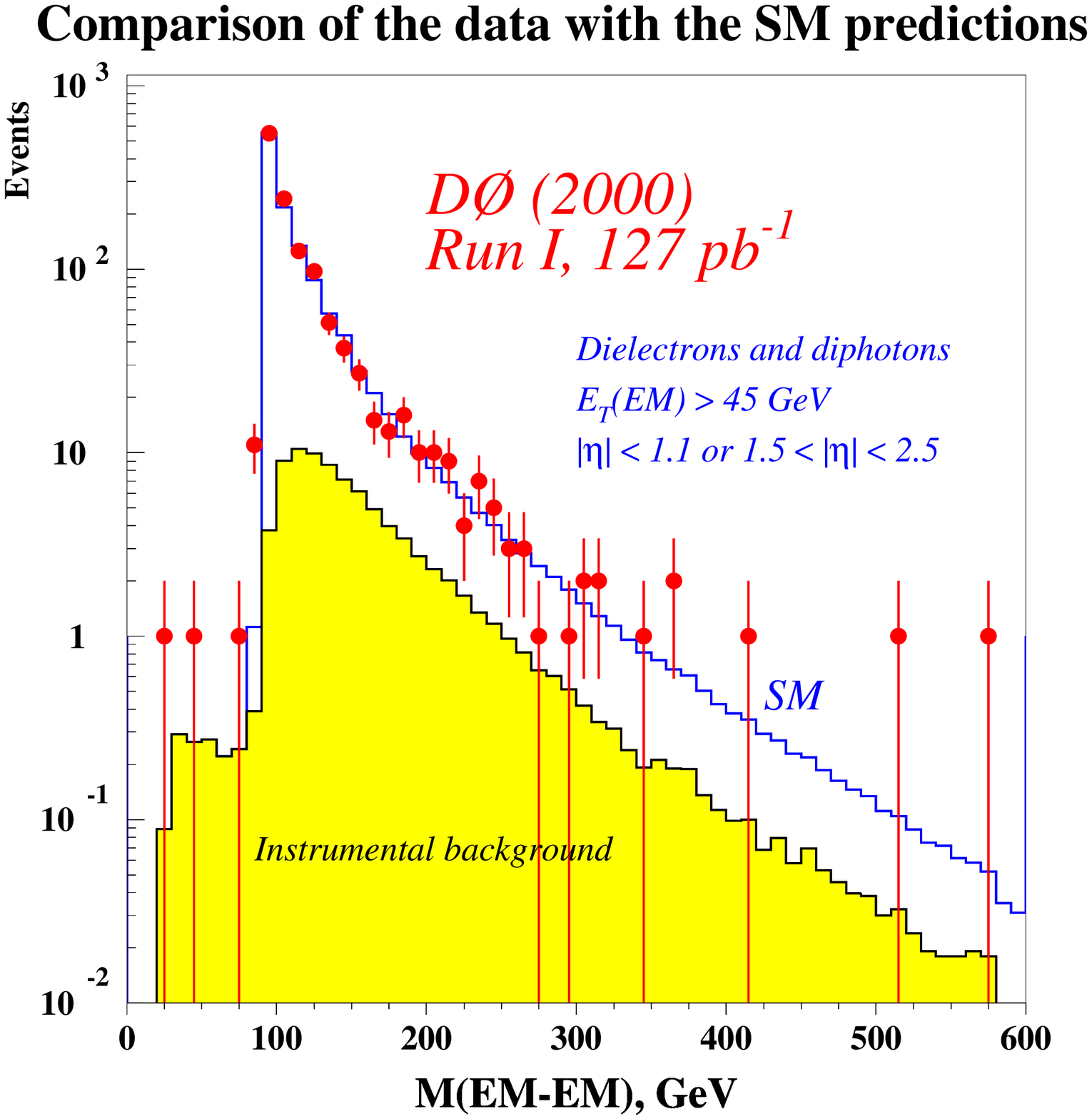}}
\caption{In the left
pane is a comparison of the decay angular distribution of the di-EM
system relative to its line of flight, and, in the right pane, of
the mass of di-EM system, with the Monte Carlo that contains
contributions only from the SM.}
\label{fig:mass_MC}
\end{figure}

Figure~\ref{fig:eta_MC} displays a comparison of data with the
Monte Carlo model.\cite{CheungL} The pseudorapidities and the
$E_T$ of the two EM objects are in excellent agreement with
expectations from the SM. A two-dimensional comparison between
data and Monte Carlo is given in Fig.~\ref{fig:two_dim} in the
($M_{_{EM,EM}},|\cos\theta^*|$) plane, and shows again the good
agreement with expectations. Projections onto the $M_{_{EM,EM}}$ and
cos$\theta^*$ axes are given in Fig~\ref{fig:mass_MC}, and confirm
the qualitative features displayed in Fig.~\ref{fig:two_dim}.

\begin{figure}
\centerline{
 \epsfxsize=4.5truein \epsfbox{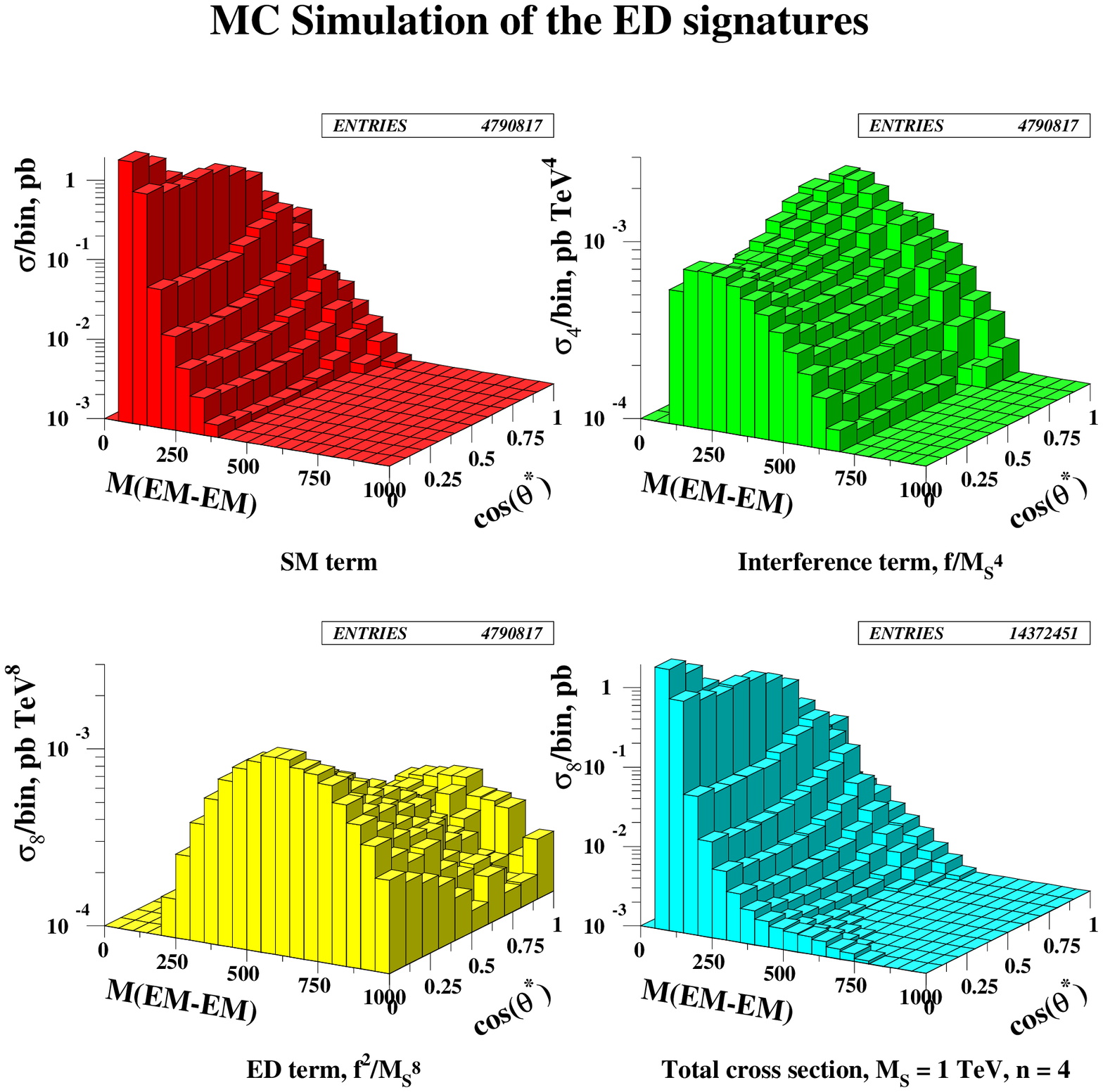}}
\caption{Contributions from gravity and all SM processes to the Monte Carlo
in the ($M_{_{EM,EM}},|\cos\theta^*|$) plane, for $M_S$= 1 TeV, and $n=4$}
\label{fig:two_dim_grav}
\end{figure}

With no excess apparent beyond expectations of the SM, D\O\ proceeds
to calculate a lower limit on the graviton contribution
to the di-EM cross section. The cross section as a function of
$M_{_{EM,EM}}$ and $|\cos\theta^*|$ can be written as:

$$\sigma = \sigma(SM) + \eta \times \sigma_4 + \eta^2 \times \sigma_8$$

\noindent where $\eta = F/{M_S}^4$, and $F=2/(n-2),$  for $n > 2$,
and where $\sigma(SM)$ represents the cross section for the SM,
$\sigma_8$ the pure graviton contribution, and the term linear in
$\eta $ is the interference between the two. (The Monte Carlo does
not include the $gg \to \gamma\gamma$ source of the SM, because it
is quite small at large $M_{_{EM,EM}}$.) The three contributions
to the Monte Carlo\cite{CheungL} are displayed in
Fig.~\ref{fig:two_dim_grav}, along with the impact on the total
$\sigma$ of a contribution from $G_{KK}$ for $M_S$= 1 TeV and
$n=4$. It is clear that the addition of $G_{KK}$ increases the
yield at large $M_{_{EM,EM}}$, especially for small values of
$|\cos\theta^*|$.

The expected sensitivity to $\eta$ is obtained from a fit of the above
formula for the cross section ($\sigma$) to Monte Carlo samples that do not
contain graviton components, and have statistics appropriate to the D\O\ data,
which corresponds to an integrated luminosity of 127 events/pb. Such fits
yield an expected sensitivity of $\eta < 0.44$ TeV$^{-4}$. The D\O\ fits are
performed using a Bayesian formalism that yields the likelihood for $\eta $,
and the expected limit of $0.44$ TeV$^{-4}$ corresponds to an upper limit
at 95\% confidence. (The procedure is also checked using a frequentist
maximum likelihood.)

The result of a similar fit to the data rather than to Monte
Carlo, yields an upper limit of $0.46$ TeV$^{-4}$, which provides
a lower limit on the value of $M_S$ that depends somewhat on the
value of $n$. In particular, $M_S > 1.44$ TeV for $n=3$, and $M_S
> 0.97$ TeV for $n=7$. The case for $n=2$ is more
delicate,\cite{PRL} but not of great interest, considering our
previous remarks on this issue.

\section{Summary}

In summary, D\O\ has presented first results of a search for contributions
of virtual gravitons to production processes at the Tevatron. Based on an
analysis of massive e-pairs and di-photons, in the context of the ADD
scenario of large extra dimensions and a single mass scale in the domain
of particle interactions, that mass scale must be greater than 1.0--1.5
TeV. These 95\% confidence limits are comparable to the final results
anticipated from LEP. Because of the ambiguities inherent in the definition
of $M_S$, e.g., from the cut-off required on the effective theory, it is
important to
check the scales under different conditions. More studies are forthcoming
from CDF and D0 on real graviton emission (mono-jet events), as well as on
virtual graviton exchange, which by the end of Run-2, should be sensitive
to scales of 3-4 TeV. Beyond that lies the LHC, with sensitivity up to 10
TeV,\cite{CheungL} beyond which the ADD idea, if not confirmed, will likely
lose much of its intriguing appeal.

\section{Acknowledgements}

I thank Greg Landsberg and my colleagues at D\O\ for providing me
the material I presented at Cairo, and also Landsberg for
his comments on this manuscript. Finally, I wish to thank Shaaban
Khalil for organizing this conference at such a wonderful
location, and for allowing speakers from the Tevatron to infiltrate the
program.


\medskip








\begin{thebibliography}{99}







\bibitem{Arkani} N. Arkani-Hamed, S. Dimopoulos, G. Dvali, Phys. Lett.
{\bf B429}, 263 (1998); I. Antoniadis, et al, Phys. Lett. {\bf B436}, 257
(1998); N. Arkani-Hamed, S. Dimopoulos, G. Dvali, Phys. Rev {\bf D59}, 086004
(1999).
\bibitem{Adelberger} E. Adelberger, Bull. Am. Phys. Soc. B3.004 (April 2000);
C. D. Hoyle, et al, HEP-PH/0011014 (Nov. 2000).
\bibitem{Kaluza} Th. Kaluza, Sitz. Preuss. Akad. Wiss. Phys. Math. Klasse,
996 (1921); O. Klein, Z. Phys. {\bf 37}, 895 (1926).
\bibitem{LEP} M. Acciarri, et al, Phys. Lett.  {\bf B464}, 135 (1999);
{\it ibid}, {\bf B470}, 268 (1999); {\it ibid}, {\bf B470}, 281 (1999);
G. Abbiendi, et al, Phys. Lett.  {\bf B465}, 303 (1999); {\it ibid},
Eur. Phys. J. {\bf C18}, 253 (2000).
\bibitem{HERA} C. Adloff, et al, Phys. Lett.  {\bf B479}, 358 (2000).
\bibitem{Gerdes} Private communication from D. Gerdes.
\bibitem{PRL} B. Abbott, et al, Phys. Rev. Lett. {\bf 86}, 1156 (2001);
See also, G. Landsberg, HEP-EX/0009038, Proc. of International
Conference on High Energy Physics at Osaka (2000).
\bibitem{Hewett} J. L. Hewett, Phys. Rev. Lett.  {\bf 82}, 4765 (1999).
\bibitem{Giudice} G. Giudice, R. Rattazzi and J. Wells, Nucl. Phys.
{\bf B544}, 3 (1999).
\bibitem{Han} T. Han, J. D. Lykken, and R. J. Zhang, Phys. Rev. {\bf D59},
105006 (1999).
\bibitem{CheungL} K. Cheung and G. Landsberg, Phys. Rev.  {\bf D62} 076003
(2000).
\bibitem{GregM} G. Landsberg and K. T. Matchev, Phys. Rev. {\bf D62},
035004 (2000).
\bibitem{Dylan} B. Abbott, et al, Phys. Rev. Lett. {\bf 84} 2792 (2000).
\end{thebibliography}
\end{document}